# Efficient and affordable thermomagnetic materials for harvesting low grade waste heat

*Daniel Dzekan[1,2], Anja Waske[3], Kornelius Nielsch[1,2], Sebastian Fähler[1]\**

[1] Leibniz IFW Dresden, Dresden, Germany

[2] Technical University of Dresden, Institute of Materials Science, Dresden, Germany.

[3] Bundesanstalt für Materialforschung und –prüfung (BAM), Berlin, Germany

*e-mail: s.faehler@ifw-dresden.de

## Abstract

*Industrial processes release substantial quantities of waste heat, which can be harvested to generate electricity. At present, the conversion of low grade waste heat to electricity relies solely on thermoelectric materials, but such materials are expensive and have low thermodynamic efficiencies. Although thermomagnetic materials may offer a promising alternative, their performance remains to be evaluated, thereby hindering their real-world application. Here we evaluate the efficiency and cost effectiveness of thermomagnetic materials that can be used in motors, oscillators and generators for converting waste heat to electricity. Our analysis reveals that up to temperature differences of several 10 K the best thermomagnetic materials outperform thermoelectric materials. Importantly, we find that the price per watt for some thermomagnetic materials are much lower compared to present-day thermoelectrics and can become competitive with conventional power plants. The materials library that we present here enables the selection of the best available thermomagnetic materials for harvesting waste heat and gives guidelines for their future development.*

Humanity finds itself at a tipping point at which the efficient use of primary energy has become decisive [1]. An important option in this regard is to recover waste heat, which is released during industrial and chemical processes in a quantity that is equivalent to almost 72 % of all electrical energy produced in the year 2016 [2]. However, most of the waste heat is just above room temperature [3], at which few existing technologies can convert heat to electricity. Although thermoelectric generators can work in this temperature range, they suffer from a low thermodynamic efficiency and high price [4, 5]. Thus, there is a strong need for alternative energy materials for the conversion of low-grade waste heat to electricity.

Magnetic materials have an outstanding position within the class of energy materials. The combination of hard and soft magnetic materials in electric motors and generators are state of art for the conversion of electrical energy to mechanical energy and vice versa. Magnetocaloric materials expand the application range further towards the conversion of electrical energy to thermal energy [6, 7]. In these materials, a steep change of magnetisation around room temperature yields a magnetically induced entropy change, which drives a highly efficient magnetocaloric cooling cycle [8]. Intense research on magnetocaloric materials [9] has led to the development of several devices and prototypes [10] that enable energy efficient cooling. The rapid progress in magnetocaloric materials has also opened up the possibility of the reverse process: converting thermal energy to electrical energy in thermomagnetic systems. Although the first concepts for thermomagnetic energy harvesting had been already suggested by Tesla [11, 12], Stefan [13], and Edison [14, 15] more than 100 years ago, it required the development of magnetocaloric materials to build the first thermomagnetic demonstrators [16].

In this paper we introduce thermomagnetic materials (TMM) as dedicated energy materials for harvesting waste heat. To identify the differences to magnetocaloric materials, we analyse the thermomagnetic harvesting cycle and describe how this cycle is implemented within thermomagnetic motors, oscillators and generators. From this, we derive generalised criteria for the magnetic and thermal properties required for optimum thermodynamic efficiency and cost effectiveness. We use these criteria to benchmark thermomagnetic materials in two Ashby plots as figures of merit and predict guidelines for their development. From our materials library we analyse the four most promising TMMs and specify their application areas. For harvesting low grade waste heat we identify La-Fe-Co-Si as the best TMM available today, which outperforms thermoelectric materials with respect to both, thermodynamic efficiency and cost effectiveness.

## The thermomagnetic cycle for harvesting waste heat

To identify the requirements for high thermodynamic efficiency of thermomagnetic materials (TMM), we first describe a thermomagnetic cycle and then identify the differences compared to a magnetocaloric cooling cycle. This generalised approach allows treating all thermomagnetic harvesting devices together, analysed in the next section.

A four step thermomagnetic cycle (Fig. 1 a) uses TMM as functional material, which reduces its magnetisation $M$ at a transition temperature $T_t$. Step I starts at ambient temperature, where the TMM is below $T_t$ and exhibits a high $M_{cold}$. At constant temperature a magnetic field $H$ is applied, which reduces the Gibbs energy $E_M$ of the TMM by $-\mu_0 M_{cold} H$, where $\mu_0$ is the magnetic field constant [17]. In step II, low grade waste heat $Q_{in}$ is used to heat the TMM above $T_t$, which reduces magnetisation to $M_{hot}$. When the magnetic field is removed in step III, just a low value of Gibbs energy term $+\mu_0 M_{hot} H$ is required. In step IV, the hot TMM is brought again into contact with ambient temperature, which closes the thermomagnetic cycle and restores the high $M_{cold}$. The difference in Gibbs energy $-\mu_0 \Delta M H$ is used to create electricity, with $\Delta M = M_{cold} - M_{hot}$ being the decisive material property. As this contribution to Gibbs energy only contains magnetic properties, we call the positive counterpart, which is harvested by thermomagnetic systems, as magnetic energy (density)

$$E_M = +\mu_0 \Delta M H \qquad (1)$$

and drop the term density for better readability. During one cycle the TMM converts $E_M$, thus a thermomagnetic system can at best convert $E_M$ to electrical energy.

A thermomagnetic cycle differs from a magnetocaloric one in two aspects. First, a thermomagnetic cycle consists of two isothermal and two isofield steps, whereas in a magnetocaloric cycle the isothermal steps are replaced by adiabatic ones [8]. Second, $Q_{in}$ is the thermal input energy and $E_M$ is the output energy, which is vice versa within a magnetocaloric cycle. Both aspects have impact on efficiency and are treated in detail later.

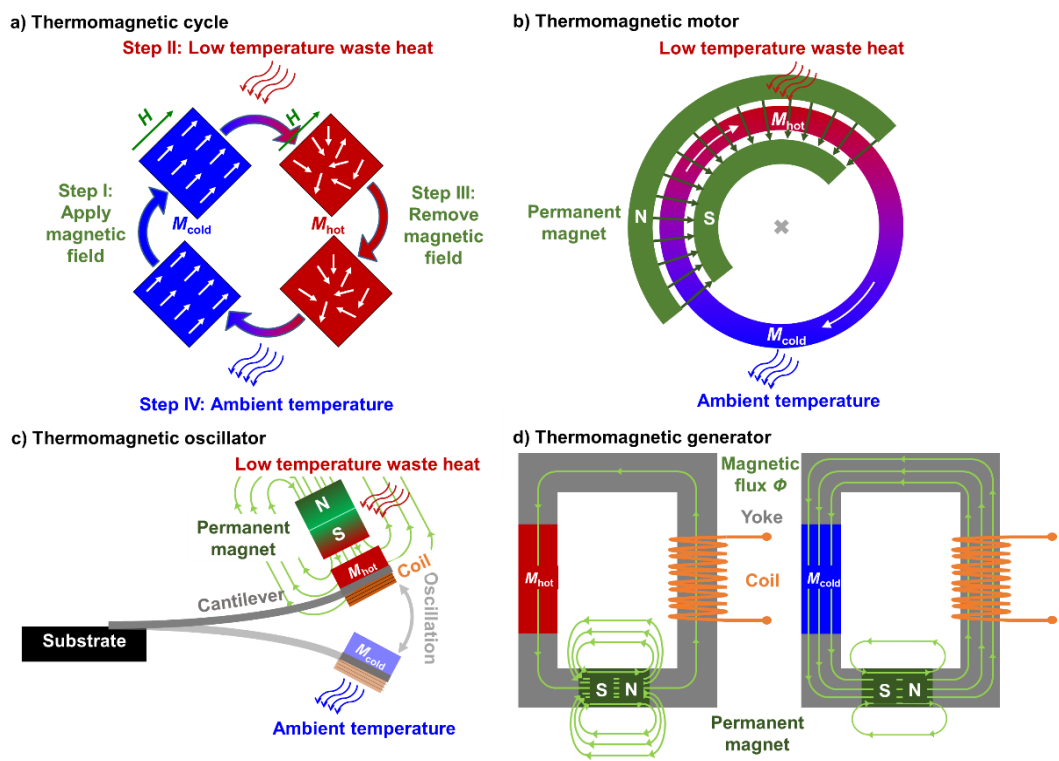

*Fig. 1. Thermomagnetic harvesting of low-temperature waste heat. a) A thermomagnetic material (TMM) is subjected to four steps within a thermomagnetic cycle. In step I, a magnetic field H is applied to the cold TMM (blue) that has a high magnetisation $M_{cold}$, which increases its Gibbs Energy. In step II, low grade waste heat $Q_{in}$ is used to increase the temperature of the TMM (red), which reduces its magnetisation to $M_{hot}$. Thus, when the magnetic field is removed in step III, there is just a low increase in Gibbs energy. In step IV, the TMM is cooled to ambient, which*

*restores its high magnetisation $M_{cold}$ and closes the thermomagnetic cycle. The difference in Gibbs energy can be harvested by one of the following setups. b) A rotatable ring of TMM (blue-red colour gradient according to its temperature) in a thermomagnetic motor, subjected to the thermomagnetic cycle. c) A TMM film is mounted at the tip of a cantilever within a thermomagnetic oscillator. An additional induction coil at the tip of the cantilever converts the mechanical oscillation within the gradient of the permanent magnet to electric energy. d) The TMM within a thermomagnetic generator is used to switch the magnetic flux Φ (green arrows) within a magnetic circuit during cycling between hot (left) and cold (right). In this circuit, Φ is created by a permanent magnet (green) and guided by a soft magnetic yoke (grey). As the magnetic field acting on the TMM changes with the flux, a thermomagnetic generator is thus an implementation of a thermomagnetic cycle.*

## Thermomagnetic motors, oscillators and generators

The thermodynamic cycle is implemented in several thermomagnetic devices, and we introduce a classification depending on the type of mechanical movement involved. As a comprehensive collection on devices had recently been given by Kishore and Priya [18], here we show that despite these different movements the TMM is always subjected to the same thermomagnetic cycle.

Mechanical rotation is employed within a thermomagnetic motor (Fig. 1b). First proposals of such devices were made by Edison [14], Tesla [11] and Stefan [13], and later works predicted the efficiency of such a device to reach the thermodynamic limit [19, 20, 21]. A thermomagnetic motor uses a rotatable ring of TMM. Its rotation causes each part of the TMM to undergo the four stages of the thermodynamic cycle. In stage I, the application of a magnetic field $H$ is realised by a permanent magnet. As the TMM exhibits a high $M_{cold}$, it is strongly attracted by the field gradient at the edge of the permanent magnet. The integral gain of mechanical energy associated with this torque is identical to the gain of $\mu_0 M_{cold} H$. In stage II, the TMM is heated by the low grade waste heat, which reduces the magnetisation to $M_{hot}$. Thus, when the TMM leaves the permanent magnet region in stage III, only a low torque hinders the rotation of the TMM ring. In step IV, the temperature of the TMM reduces to ambient and restores the high $M_{cold}$. Thus, in a thermomagnetic motor the heat $Q_{in}$ is used to convert the magnetic energy $E_M = \mu_0 \Delta M H$ into mechanical energy, which can be converted to electrical energy by a conventional generator. Although many miniature versions of this motor, known also as "Curie wheel", can be viewed on video-sharing websites, these motors can also be more powerful: e.g. a prototype using gadolinium as the TMM reached a power of 1.4 kW [22].

Mechanical oscillation is used within thermomagnetic microsystems. In these systems, the TMM is used in the shape of a thin film deposited on top of a vibrating cantilever (Fig. 1c). In step I of the thermomagnetic cycle, the cold TMM film is attracted by a permanent magnet, which bends the cantilever. The permanent magnet is combined with the heat source, and thus in step II the temperature of the TMM increases, which reduces its magnetisation. This decreases the attractive force of the TMM towards the permanent magnet. Accordingly, in step III, the restoring force of the bent cantilever is sufficient to move the TMM away from the heat source. With the heat source also being the permanent magnet, $H$ is reduced. At sufficient distance, the TMM cools to ambient (step IV). The mechanical energy of the vibrating cantilever is converted to electrical energy by an induction coil, which is located on top of the cantilever [23]. During vibration, this coil moves within the magnetic field gradient of the permanent magnet and thus, according to Faraday's law of induction, the flux change induces an electric voltage. In a different design [24], as suggested by the group of Carman [25, 26, 27], a piezoelectric cantilever is used instead of the coil. Although bulk thermomagnetic oscillators, which act like a linear motor, have also been demonstrated [28], the advantage of a microsystem is its fast heat exchange, which is possible because of the reduced size of the TMM. This results in a high frequency of the thermomagnetic cycle, which can reach up to 200 Hz when the resonance frequency of the cantilever matches the thermal exchange frequency [23].

No mechanical movement of the TMM is required for a thermomagnetic generator. First concepts of generators were invented by Edison [15] and Tesla [12], and later Brillouin and Iskenderian calculated the efficiency relative to Carnot to be up to 55 % [29]. Based on this work, other researchers treated such a device theoretically [30, 31, 32, 33]. In this implementation of a thermomagnetic cycle, the TMM is used as a thermal switch for the magnetic flux Φ, which is created by a permanent magnet (Fig. 1d). At low temperatures, the high $M_{cold}$ of the TMM guides Φ through a closed magnetic circuit. At high temperatures, the low $M_{hot}$ opens the magnetic circuit and reduces Φ. Following Faraday's law of induction, this flux change induces an electric voltage. To harvest electric energy, an induction coil is wound around the soft magnetic yoke, connecting the permanent magnet and TMM. The flux change during step II and IV converts magnetic energy $E_M$ into electrical energy. Opening and closing the magnetic circuit also changes the magnetic field H that acts on the TMM, as illustrated by the different density of flux lines in Fig. 1d. The first proof of concept was published by Srivastava et al. in 2011 [16]. The efficiency of this demonstrator was quite low, mainly because of an unoptimised magnetic circuit. As a large difference in Φ is beneficial to increase $E_M$, more complex magnetic field topologies have been used for thermomagnetic generators. A topology with two magnetic circuits avoids magnetic stray fields [34]. A topology with three circuits even allows a sign reversal of Φ, which increases both, output voltage and power, by orders of magnitude [35].

## Efficiency of thermomagnetic materials

Thermodynamic efficiency is a key property of each energy harvesting process as it defines the fraction of usable output energy vs. thermal input energy $Q_{in}$ during each cycle. For TMM the magnetic energy $E_M$ is the upper limit for the output energy, which gives the material efficiency:

$$\eta = \frac{E_M}{Q_{in}} = \frac{\mu_0 \Delta M H}{Q_{in}} \quad (2)$$

Most implementations today reach much lower values for the system efficiency [34, 35]. Possible reasons for this are thermodynamic cycles that do not consist of strict isothermal/isofield steps, losses by insufficient insulation, which are challenging at low frequencies, or an incomplete use of $Q_{in}$ to heat up the TMM. Here we focus on the upper limits as defined by material efficiency, and do not address the system engineering aspects as we expect a strong improvement from the very few existing prototypes.

For many TMM the physical quantities used in Eq. (2) are available and summarised in Supplementary Table S1, since they can be extracted from available measurements, as described in Supplementary Fig. S2. The applied magnetic field $H$ is the only property in the equation that depends on the device and not on the material. To compare material properties, we fix $\mu_0 H = 1$ T, given that a field of 1 T can be easily obtained by currently available permanent magnets. As minimum field we used zero, which can be approached well by magnetic shielding as used e. g. in magnetocaloric devices [36].

To identify the TMM with highest $\eta$, we evaluated the magnetisation change $\Delta M$ and heat input $Q_{in}$ for several materials and summarise them in Fig. 2 in Ashby type plots. We selected only the materials with a transition temperature between 273 K and 373 K, at which water can be used as a heat-transfer fluid. This transition temperature can be tuned in many material systems by adjusting the composition, which can be used to adapt $T_t$ to the available waste heat. Pure magnetic elements (Fe, Ni, Co) are shown only for comparison, as they have $T_t$ far above the relevant temperature range. The benchmark is the Carnot efficiency $\eta_{carnot} = \Delta T/T_{hot}$, which represents the upper theoretical limit according to thermodynamics.

The most efficient TMM are ones in which a maximum of $\Delta M$ is obtained at a minimum $Q_{in}$: these are found in the top left corner of the Ashby plots. We selected two different values for $\Delta T$: 10 K and 30 K. For the low value of ΔT, materials with the tendency of a first order transition are the most suitable as they exhibit a sharp transition where $\Delta M$ changes in a narrow ΔT. For a large ΔT, materials with a second order transition become competitive, which exhibit a gradual decrease in magnetisation during heating. The best TMM can reach an efficiency of about 2 %, which is about 60 % of the upper limit of $\eta_{carnot}$ = 3.3 % for $T_{hot}$ = 300 K and $\Delta T$ = 10 K. As we will analyse in detail later, this illustrates that even the existing TMM can compete with thermoelectrics. This is quite astonishing since most of the materials analysed here had been developed for magnetocaloric refrigeration, which must convert a large amount of heat during each cycle. For energy harvesting a new paradigm for materials development is necessary: TMM should consume as low as possible $Q_{in}$ and accordingly one must aim for low heat capacity and latent heat.

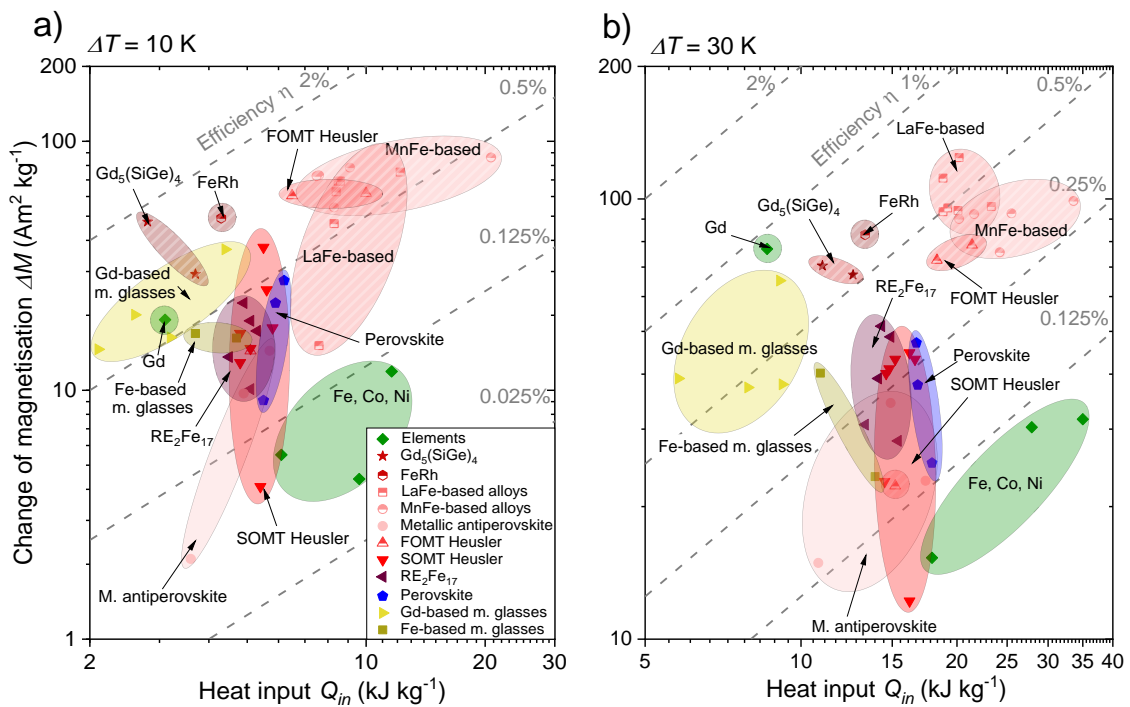

***Fig. 2. Evaluating the thermodynamic efficiency η of thermomagnetic materials.** To reach high η, a large change of magnetisation ΔM is beneficial, as well as a low heat input $Q_{in}$. The grey dashed lines represent a constant efficiency $\eta = \mu_0 \Delta M H Q_{in}^{-1}$. Accordingly the most efficient materials are located in the top left corner, where η approaches 2 %. Material properties were evaluated for two different temperature spans in Ashby-type plots. **(a)** At ΔT = 10 K materials exhibiting a first order transition (half solid symbols) reach the highest efficiencies. **(b)** At ΔT = 30 K materials with a*

*second order transition (solid symbols) become competitive. Metallic materials are displayed in shades of red, ceramics in blue, metallic glasses in shades of yellow, and elements in green.*

## Power density, specific cost and economic feasibility

A decisive criterion which TMM will be used for real-world applications is its cost effectiveness, which describes the price for each watt of electric power harvested. To probe if cost effectiveness of TMM can compete with thermoelectrics or even conventional sources of electrical energy, we expand our analysis of thermodynamic efficiency towards power density and specific cost.

The power density $P_D$ describes the power per unit volume that can be harvested by a TMM: $P_D = E_{\text{mag}} \cdot f$. High $P_D$ requires high energy per cycle $E_M$ as well as a high cycle frequency $f$. The latter requires fast heat exchange which, in turn, relies on high thermal conductivity $\lambda$ and a low volumetric specific heat $\rho c_p$, where $\rho$ is the density. We used a one-dimensional lumped capacitance method to derive an analytical formula for $f$ (see Methods section for details). In addition to the materials thermal properties (Supplementary Fig. S3), only the thickness $d$ of the TMM is required. We selected $d$ = 0.5 mm, because this thickness is achievable by most bulk-processing routes. Indeed, for the La-Fe-Co-Si materials, plates with this thickness are already available commercially.

To estimate the specific material cost $C$ we used the current raw-material prices per volume. We did not include the costs of processing and shaping in the present analysis, as these costs depend on the scale of production and are expected to decrease strongly once thermomagnetic harvesting is established. Furthermore, we considered only the TMM and not the periphery required for a complete thermomagnetic system (hard magnets, soft magnetic yoke, tubing, etc.). Our cost estimates thus represent the lower limit. A realistic estimate of the periphery is not possible at the present level of technology readiness, but note that the use of cheap ferrite hard magnets appears possible [35], whereas magnetocaloric refrigeration requires expensive Nd-Fe-B.

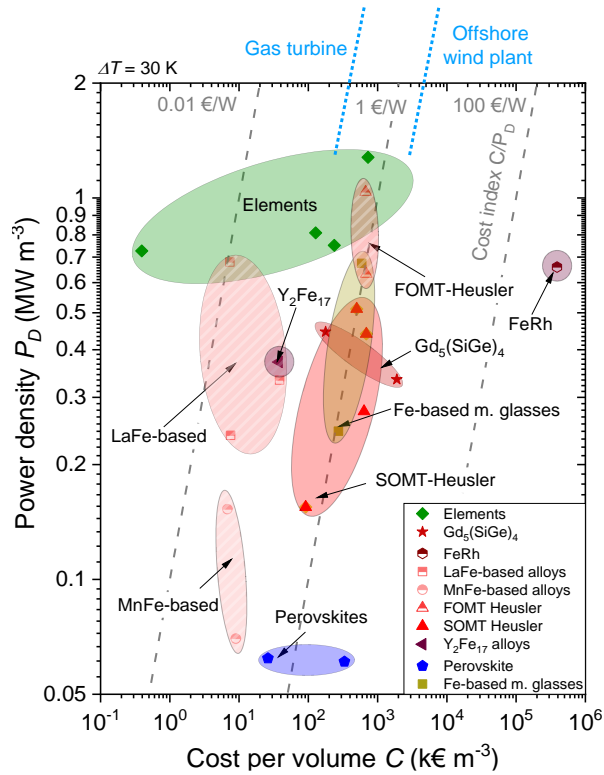

***Fig. 3. Identifying the most cost-effective thermomagnetic materials.*** *Ashby-type plot of power density versus cost per volume for ΔT = 30 K. The diagonal grey lines depict the cost index $C/P_D$ and accordingly the most cost-effective materials are found in the top-left edge. For comparison the levelised costs of current energy technologies are noted at the top. Though C considers only the cost of the active material and not the periphery (yoke, permanent magnets, processing, etc.), this illustrates that harvesting low grade waste by thermomagnetic material can become competitive with conventional power plants as no additional primary energy is needed.*

To evaluate the cost effectiveness of TMM, we calculated the cost index $C/P_D$, which gives the price in Euro required for each Watt of output power (Fig. 3). This allows a rough comparison with the costs of common power generation, that is quantified by the Levelised Costs Of Electricity (LCOE), which considers the construction, operation and financing of a power plant during life time. The LCOE of present-day power plants range from 0.4 €/W for gas turbines to 4 €/W for offshore wind plants [37].

Thermoelectric power generation requires about 25 €/W, considering only the material, which is obviously worth it for specific applications when no power grid is available [4]. The pure metals (green) in particular iron in Fig. 3 reach the lowest cost index, but are not suitable due to their low thermodynamic efficiency (Fig. 2). Among the materials with high efficiency the lowest cost index is obtained for La-Fe based TMM. Its cost index is more than one order of magnitude lower than the LCOE of present-day power plants. This should leave a sufficient budget for building a complete thermomagnetic system, not least given that waste heat is available for free. We expect that the high cost effectiveness of TMM will be more decisive for the success of thermomagnetic harvesting than thermodynamic efficiency, which describes only how much waste heat is required.

# Optimum thermomagnetic materials today and guidelines for future development

The Ashby plots (Fig. 2 and Fig. 3) helped us identify the best classes of TMM that are currently available in terms of their thermodynamic efficiency and cost effectiveness. Here we focus on the four optimum alloy compositions and discuss their suitability for particular applications. Furthermore, we derive guidelines for improving TMM further by comparing the different material classes.

The properties of the four most promising TMM and their particular compositions are summarised in Fig. 4. Gadolinium, which is still the benchmark for magnetocaloric refrigeration [38], is added as reference. Whereas the first order $LaFe_{11.8}Si_{1.8}H_1$ exhibits the optimum combination of all properties for bulk applications, $Mn_{1.25}Fe_{0.7}P_{0.5}Si_{0.5}$ has a substantially lower thermal diffusivity, which reduces frequency and, accordingly, power density. The first order Heusler alloys reach the highest power density, but they exhibit a very high specific material cost. Although this cost will hinder bulk application, this is not the case with application in microsystems, for which the material cost is less important than the processing cost. $Y_2Fe_{17}$ is the most promising thermomagnetic material with a second order transition.

To identify guidelines for optimum TMM, in both Ashby plots the different material classes are colour and symbol coded. TMM with a first order transition reach higher efficiency at $\Delta T$ = 10 K, which we attribute to their steep change of magnetisation. Accordingly, some first order materials also achieve an excellent cost index. However, this only holds for La-Fe- and Mn-Fe-based materials, which both have moderate material costs due to their high content of inexpensive Fe. Furthermore, in both of these material classes the composition can be used to tune the transformation close to second order [39, 40]. By this one keeps a reasonably steep change of magnetisation, but avoids the large hysteresis, which often occurs in first order materials and hinders full reversibility [41]. At $\Delta T$ = 30 K materials with a second order transition become competitive; these materials require a larger $\Delta T$ and $Q_{in}$ to reach a sufficiently high $\Delta M$. Thus second order materials are interesting when a large temperature span of waste heat is available.

Crystalline metallic materials can reach higher values of $\Delta M$ compared to ceramics as they have a higher magnetisation because of a higher density of ferromagnetic elements. Furthermore, ceramics have a low thermal conductivity as compared to metals because of the absence of free electrons in the former, and thus their power density is low. Metallic glasses typically have even lower $\Delta M$ because of their broad second order transition, but the equally reduced $Q_{in}$ results in a competitive thermodynamic efficiency. However, additional alloying is required to enable the formation of glass, which makes glasses expensive. Thus, we propose that future work focus on crystalline metallic materials with a high content of Iron when searching for better TMM. A more detailed evaluation of each material system is provided in Supplementary Table S2.

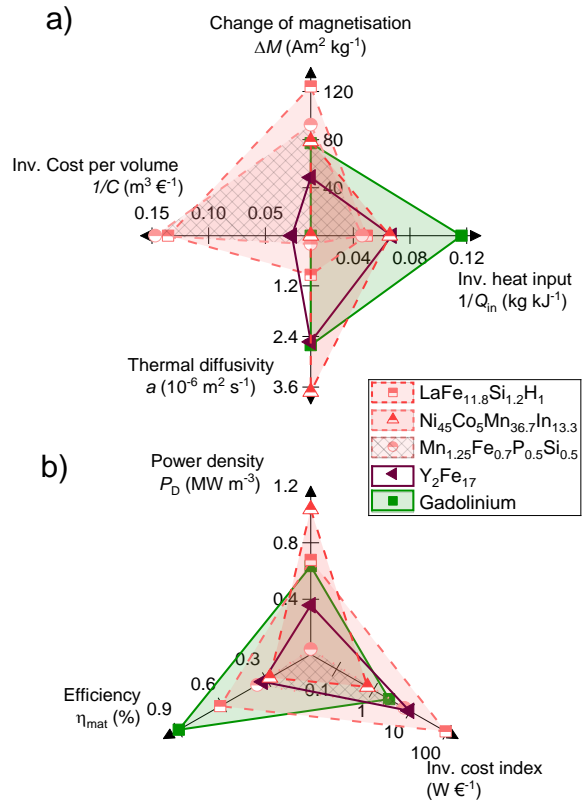

*Fig. 4. The four most promising thermomagnetic materials. a) Physical and economic properties. b) Thermomagnetic properties. The radar charts are generated by using the best composition for each material class for $\Delta T = 30\ K$. Gd is depicted as a reference.*

## Thermomagnetic vs. thermoelectric materials

Our analysis of cost effectiveness reveals that thermomagnetic can outperform thermoelectric materials by three orders of magnitude. In this section we return to thermodynamic efficiency in order to benchmark the best thermodynamic and thermoelectric materials with respect to $\Delta T$.

Following our previous analysis, La-Fe-based TMM are most promising for bulk applications. For this benchmark we selected the particular La-Fe-Co-Si system, as plates are already available commercially (Calorivac C® [42]). This geometry is favourable for most applications, as it enables both, guiding the magnetic flux within the plate and a fast perpendicular heat transfer [35]. The hydrogenised La-Fe alloys, which reach slightly higher values, are not available in plate shape. To determine $\eta(T)$ (Eq. 2), we measured Δ*M* and $Q_{in}$ in dependency of temperature (Supplementary Fig. S1). Both properties exhibit different dependencies on Δ*T*. Whereas Δ*M* has a favourable high value just at the transition temperature, $Q_{in}$ increases continuously with increasing Δ*T*, and large temperature differences are hence unfavourable. Accordingly the efficiency of thermomagnetic harvesting increases strongly at low Δ*T* (Fig. 5).

The efficiencies of two thermoelectric materials are shown for comparison. $Bi_2Te_{2.79}Se_{0.21}$ is the material reaching the highest values around room temperature [43]. The organic thermoelectric material PEDOT:PSS [44] has lower values, but this material is more suitable for this low temperature range as it is cheaper and more sustainable [5]. At temperature differences below 10 K TMM reach much higher efficiencies than thermoelectric materials, which is a key advantage for several applications. An example of such low grade waste heat is geothermal energy and thus we expect that early applications of thermomagnetic harvesting will use this source for off-grid devices. A second example is body heat, which makes thermomagnetic microsystems in particular suitable to power e. g. smartwatches [45].

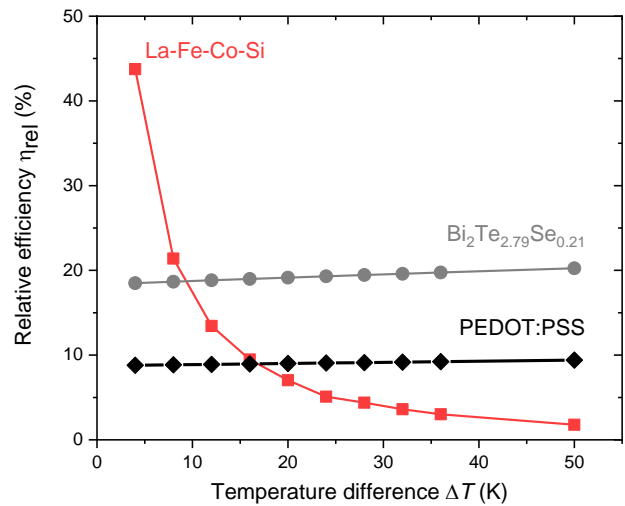

***Fig. 5. Efficiency relative to Carnot of TMM and thermoelectric materials for harvesting low-temperature waste heat.*** *The relative efficiency $\eta_{rel}$ of the TMM like La-Fe-Co-Si (Calorivac C®) increases strongly at low temperature differences. For comparison $Bi_2Te_{2.79}Se_{0.21}$ as the most efficient thermoelectric material [43] and PEDOT:PSS [44] as more sustainable organic material with lower cost are shown. The high efficiency of TMM at temperature differences up to several 10 K is of particular advantage for harvesting body temperature and geothermal energy.*

## Summary and Outlook

Thermomagnetic materials (TMM) enable the conversion of low-temperature waste heat to electricity by the change of magnetisation with temperature. Though their implementations in harvesting systems like motors, oscillators, and generators differ with respect to the mechanical motion, the TMM always follows the same thermodynamic cycle. This allows for an universal evaluation of TMM from their materials properties. We present two Ashby type charts – one for thermodynamic efficiency and another for cost effectiveness – which serve as figures of merit for TMM. These materials library enables scientists and engineers to select the optimum thermomagnetic material for their demand.

We identify several TMM that outperform even the best thermoelectric materials for temperature differences below 10 K with respect to efficiency. This is a decisive advantage, because the largest amount of waste heat is available just above room temperature. Furthermore, our analysis reveals that the price per watt of the best TMM is more than one order of magnitude lower than that of established power technologies and three orders of magnitude lower than thermoelectric materials. This price is competitive enough to make it economically feasible to realise a complete thermomagnetic system.

The optimum combination of high thermodynamic efficiency and high cost effectiveness is currently obtained by La-Fe-based TMM. However, this material system was developed for magnetocaloric cooling applications and not for converting low-temperature heat to electricity. For developing even better TMM we propose a new paradigm for materials development. While magnetocaloric materials aim for converting a large amount of heat during each cycle, TMM should consume as little heat as possible. We predict that the next-generation TMM will be metallic, crystalline, contain a high amount of iron, and exhibit a transition at the border between first and second order.

Our materials library establishes TMM as independent class of emerging energy materials, which has the potential to become a game changer for recovering low grade waste heat due to their high thermodynamic efficiency and cost effectiveness.

**Methods**

**Experimental methods.** Temperature- and magnetic-field dependent magnetic properties of the La-Fe-Co-Si alloy (Calorivac C®[42]) with $T_t$ = 308 K were measured in a Quantum Design PPMS using a vibrating sample magnetometer insert. The heat capacity was measured in Quantum Design PPMS with heat capacity option. For all temperature-dependent properties a heating and cooling rate of 0.5 K min⁻¹ was used to avoid thermal lag between the device and the sample.

**Theoretical methods.** The data on materials were digitised from their primary sources. The magnetisation difference was derived from temperature-dependent measurements at sufficient magnetic fields or from field-dependent measurements at 1 T. The heat-capacity data are from zero-field measurements. The heat capacity was calculated as mean value $c_p = \int_{T_{cold}}^{T_{hot}} c_p \mathrm{d}T / \Delta T$ for a temperature span of 30 K. The digitised data were integrated to calculate the magnetic energy and the heat input. The temperature-dependent heat conductivity and mass density were taken in the vicinity of the transition temperature. Raw material costs were compiled from different material marketplaces in June 2018. All data are listed in Supplementary Table S1.

The power density of the TMM was calculated as the product of magnetic energy per cycle and the cycle frequency. The cycle frequency is determined as the time to heat the material up and cool it, respectively. To approximate the time in this transient conduction problem, the one dimensional lumped capacitance method (LCM) was used [46]. Within this model the solid is spatially uniform and temperature gradients are negligible. When the temperature gradient approaches zero, the heat conductivity has to be infinitive following Fourier's law. Accordingly, an overall energy balance rather than the heat equation is used to determine the transient temperature response. Thus the time $t$ required for a solid to reach a given temperature $T$ can be written as $t = \frac{\rho c_p L_c^2}{\lambda \cdot Bi} \ln \frac{T_i - T_\infty}{T - T_\infty}$, whereby $T_\infty$ is the end and $T_i$ the start temperature. The Biot number $Bi$ describes the ratio of thermal resistance of conduction to convection $Bi = \frac{R_{\text{cond}}}{R_{\text{conv}}} = \frac{hL_c}{\lambda}$. Thus, the Biot number is defined by heat transfer coefficient $h$, the heat conductivity $\lambda$ and the characteristic length $L_c$ of the solid. This length could be either the half plate thickness or the volume to surface ratio. The LCM is only valid when the temperature gradient within the solid is negligible. This is the case when resistance to conduction within the solid is much less than the resistance to convection and thus with *Bi* ≪ 1. This is fulfilled for very small characteristic length or high thermal conductivities. For the calculations of the heat transfer, the Biot number was set to be 0.1. Additionally, the term $\ln \frac{T_i - T_\infty}{T - T_\infty} = 1$ was set, which means that about 63 % of the end temperature (270 K for cooling) is reached. Supplementary Fig. S7 shows that this value results in the maximal power output, although the magnetisation change is reduced slightly. Accordingly, the cycle frequency of heating and cooling is calculated as $f = \frac{0.1 \cdot \lambda}{2 \cdot c_p \, \rho L_c^2}$.

The efficiency η of thermoelectric materials (Fig. 5) was calculated using $\eta = \frac{T_{hot} - T_{cold}}{T_{hot}} \cdot \frac{\sqrt{1 + Z\overline{T}} - 1}{\sqrt{1 + Z\overline{T}} + T_{cold}/T_{hot}}$, where ZT is the unitless figure of merit at the application temperature and $\overline{T}$ is the average temperature of the hot $T_{hot}$ and cold temperature $T_{cold}$ [43].

**Availability of computer code and algorithm, data and material**

All data used for the comparison of TMM and the references to their origin are available in Supplementary Table S1.

**Author contributions**

D.D. compared the different TMM and wrote the first version of this paper. A.W. contributed to the extrinsic properties and comparison with magnetocaloric materials. K.N. supervised the thesis of D.D. and contributed to the comparison with thermoelectric materials. S.F. suggested to make this analysis and wrote the second version of this paper. All authors contributed to the final version.

**Competing interests**

We declare no competing interests.

**Acknowledgment**

This work was funded by the German Research Foundation (DFG) by project FA453/14. The authors acknowledge M. Kohl, D. Kamble, A. Diestel, L. Fink, and D. Berger for helpful discussions.